\documentclass[preprint]{aastex631}

\newcommand*\patchAmsMathEnvironmentForLineno[1]{%
\expandafter\let\csname old#1\expandafter\endcsname\csname #1\endcsname
  \expandafter\let\csname oldend#1\expandafter\endcsname\csname end#1\endcsname
  \renewenvironment{#1}%
     {\linenomath\csname old#1\endcsname}%
     {\csname oldend#1\endcsname\endlinenomath}}%
\newcommand*\patchBothAmsMathEnvironmentsForLineno[1]{%
  \patchAmsMathEnvironmentForLineno{#1}%
  \patchAmsMathEnvironmentForLineno{#1*}}%
\AtBeginDocument{%
\patchBothAmsMathEnvironmentsForLineno{equation}%
\patchBothAmsMathEnvironmentsForLineno{align}%
\patchBothAmsMathEnvironmentsForLineno{flalign}%
\patchBothAmsMathEnvironmentsForLineno{alignat}%
\patchBothAmsMathEnvironmentsForLineno{gather}%
\patchBothAmsMathEnvironmentsForLineno{multline}%
}

\usepackage{textcomp}  
\usepackage{comment}
\usepackage{subfigure}
\usepackage{physics}
\usepackage{color}

\shorttitle{SnSearchKam2022}
\shortauthors{KamLAND collaboration}

\graphicspath{{./}{figures/}}

\begin{document}

\title{Search for supernova neutrinos and constraint on the galactic star formation rate with the KamLAND data}

\correspondingauthor{M.~Eizuka}
\email{minori@awa.tohoku.ac.jp}

\newcommand{\tohoku}{\affiliation{Research Center for Neutrino Science, Tohoku University, Sendai 980-8578, Japan}}
\newcommand{\tohokuRigaku}{\affiliation{Department of Physics, Tohoku University, Sendai 980-8578, Japan}}
\newcommand{\fris}{\affiliation{Frontier Research Institute for Interdisciplinary Sciences, Tohoku University, Sendai, 980-8578, Japan}}
\newcommand{\gppu}{\affiliation{Graduate Program on Physics for the Universe, Tohoku University, Sendai 980-8578, Japan}}
\newcommand{\ipmu}{\affiliation{Institute for the Physics and Mathematics  of the Universe, The University of Tokyo, Kashiwa 277-8568, Japan}}
\newcommand{\tokyo}{\affiliation{Department of Physics, Faculty of Science, the University of Tokyo, Tokyo 113-0033, Japan}}
\newcommand{\osakarcnp}{\affiliation{Graduate School of Science, Osaka University, Toyonaka, Osaka 560-0043, Japan}}   
\newcommand{\osaka}{\affiliation{Research Center for Nuclear Physics (RCNP), Osaka University, Ibaraki, Osaka 567-0047, Japan}}
\newcommand{\tokushima}{\affiliation{Department of Physics, Tokushima University, Tokushima 770-8506, Japan}}
\newcommand{\tokushimaStudent}{\affiliation{Graduate School of Advanced Technology and Science, Tokushima University, Tokushima, 770-8506, Japan}}
\newcommand{\kyoto}{\affiliation{Department of Physics, Kyoto University, Kyoto 606-8502, Japan}}
\newcommand{\lbl}{\affiliation{Nuclear Science Division, Lawrence Berkeley National Laboratory, Berkeley, CA 94720, USA}}
\newcommand{\hawaii}{\affiliation{Department of Physics and Astronomy, University of Hawaii at Manoa, Honolulu, HI 96822, USA}}
\newcommand{\mituniv}{\affiliation{Massachusetts Institute of Technology, Cambridge, MA 02139, USA}}
\newcommand{\bu}{\affiliation{Boston University, Boston, MA 02215, USA}}
\newcommand{\tennessee}{\affiliation{Department of Physics and Astronomy,  University of Tennessee, Knoxville, TN 37996, USA}}
\newcommand{\tunl}{\affiliation{Triangle Universities Nuclear Laboratory, Durham, NC 27708, USA}}    
\newcommand{\chapehill}{\affiliation{The University of North Carolina at Chapel Hill, Chapel Hill, NC 27599, USA}}

\newcommand{\northcarolina}{\affiliation{North Carolina Central University, Durham, NC 27701, USA}}
\newcommand{\duke}{\affiliation{Department of Physics at Duke University, Durham, NC 27705, USA}}
\newcommand{\seattle}{\affiliation{Center for Experimental Nuclear Physics and Astrophysics, University of Washington, Seattle, WA 98195, USA}}
\newcommand{\nikhef}{\affiliation{
Nikhef and the University of Amsterdam, Science Park, Amsterdam, The Netherlands}}
\newcommand{\virginia}{\affiliation{Center for Neutrino Physics, Virginia Polytechnic Institute and State University, Blacksburg, VA 24061, USA}}

\newcommand{\currentKozkov}{\affiliation{Present address: National Research Nuclear University ``MEPhI'' (Moscow Engineering Physics Institute), Moscow, 115409, Russia}}
\newcommand{\currentDima}{\affiliation{Present address: Department of Physics and Astronomy, University of Alabama, Tuscaloosa, Alabama 35487, USA}}

\newcommand{\currentHayashida}{\affiliation{Present address: Imperial College London, Department of Physics, Blackett Laboratory, London SW7 2AZ, UK}}
\newcommand{\currentUeshima}{\affiliation{Present address: National Institutes for Quantum and Radiological Science and Technology (QST), Hyogo 679-5148, Japan}}
\newcommand{\currentTakemoto}{\affiliation{Present address: Kamioka Observatory, Institute for Cosmic-Ray Research, The University of Tokyo, Hida, Gifu 506-1205, Japan}}

\author{S.~Abe}\tohoku
\author{S.~Asami}\tohoku
\author[0000-0002-1043-3438]{M.~Eizuka} \tohoku 
\author{S.~Futagi} \tohoku
\author{A.~Gando}\tohoku
\author{Y.~Gando}\tohoku
\author{T.~Gima}\tohoku 
\author{A.~Goto} \tohoku
\author[0000-0002-4238-7990]{T.~Hachiya}\tohoku
\author{K.~Hata} \tohoku
\author{K.~Hosokawa} \altaffiliation{Present address: Kamioka Observatory, Institute for Cosmic-Ray Research, The University of Tokyo, Hida, Gifu 506-1205, Japan} \tohoku
\author[0000-0001-9783-5781]{K.~Ichimura} \tohoku  
\author[0000-0001-7694-1921]{S.~Ieki} \tohoku
\author{H.~Ikeda}\tohoku
\author{K.~Inoue}\tohoku 
\author[0000-0001-9271-2301]{K.~Ishidoshiro}\tohoku
\author{Y.~Kamei} \tohoku
\author[0000-0003-2350-2786]{N.~Kawada} \tohoku
\author{Y.~Kishimoto} \tohoku \ipmu
\author{M.~Koga}\tohoku \ipmu
\author{M.~Kurasawa} \tohoku
\author{N.~Maemura}\tohoku
\author{T.~Mitsui}\tohoku
\author{H.~Miyake}\tohoku
\author{T.~Nakahata} \tohoku
\author{K.~Nakamura}\tohoku 
\author{K.~Nakamura}\tohoku 
\author{R.~Nakamura}\tohoku
\author{H.~Ozaki}\tohoku \gppu
\author[0000-0003-2855-6505]{T.~Sakai} \tohoku 
\author{H.~Sambonsugi}\tohoku
\author{I.~Shimizu}\tohoku
\author[0000-0002-3988-2309]{J.~Shirai}\tohoku
\author{K.~Shiraishi}\tohoku
\author{A.~Suzuki}\tohoku
\author{Y.~Suzuki}\tohoku 
\author{A.~Takeuchi} \altaffiliation{Present address: Department of Physics, Faculty of Science, the University of Tokyo, Tokyo 113-0033, Japan}\tohoku 
\author{K.~Tamae}\tohoku
\author[0000-0002-2363-5637]{H.~Watanabe}\tohoku
\author{Y.~Yoshida} \tohoku

\author[0000-0003-3488-3553]{S.~Obara} \altaffiliation{Present address: National Institutes for Quantum and Radiological Science and Technology (QST), Sendai 980-8579, Japan} \fris

\author{A.~Ichikawa} \tohokuRigaku


\author{S.~Yoshida}\osakarcnp
\author{S.~Umehara}\osaka

\author{K.~Fushimi}\tokushima
\author{K.~Kotera} \tokushimaStudent
\author{Y.~Urano} \tokushimaStudent


\author{B.~E.~Berger}\lbl \ipmu
\author[0000-0002-7001-717X]{B.~K.~Fujikawa}\lbl \ipmu
\author{J.~G.~Learned}\hawaii
\author{J.~Maricic}\hawaii
\author{S.~N.~Axani}\mituniv
\author{L.~A.~Winslow}\mituniv
\author{Z.~Fu}\mituniv
\author{J.~Smolsky}\mituniv
\author{Y.~Efremenko}\tennessee \ipmu
\author{H.~J.~Karwowski}\tunl \chapehill
\author{D.~M.~Markoff}\tunl \northcarolina
\author{W.~Tornow}\tunl \duke \ipmu
\author[0000-0002-4844-9339]{A.~Li}\chapehill 
\author{J.~A.~Detwiler}\seattle \ipmu
\author{S.~Enomoto}\seattle \ipmu
\author[0000-0002-1577-6229]{M.~P.~Decowski}\nikhef \ipmu
\author{C.~Grant}\bu
\author{H.~Song}\bu
\author{T.~O'Donnell}\virginia
\author{S.~Dell'Oro}\virginia

\collaboration{99}{(KamLAND Collaboration)}

\begin{abstract}
We present the results of a search for core-collapse supernova neutrinos, using long-term KamLAND data from 2002 March 9 to 2020 April 25.
We focus on the electron antineutrinos emitted from supernovae in the energy range of 1.8--111\,MeV.
Supernovae will make a neutrino event cluster with the duration of $\sim$10\,s in the KamLAND data.
We find no neutrino clusters and give the upper limit on the supernova rate as to be 0.15\,yr$^{-1}$ with a 90\% confidence level. 
The detectable range, which corresponds to a $>$95\% detection probability, is 40--59\,kpc and 65--81\,kpc for core-collapse supernovae and failed core-collapse supernovae, respectively. 
This paper proposes to convert the supernova rate obtained by the neutrino observation to the Galactic star formation rate.
Assuming a modified Salpeter-type initial mass function, the upper limit on the Galactic star formation rate is $<$(17.5--22.7)\,$M_{\odot}\,\mathrm{yr}^{-1}$ with a 90\% confidence level.
\end{abstract}


\section{Introduction} \label{sec:intro}
On 1987 February 23, neutrinos from SN 1987A, a supernova explosion in the Large Magellanic Cloud, were observed. 
This was the first observation of a supernova neutrino burst, and extrasolar neutrino astronomy begun at this time. 
Even though the number of detected events was only 24~\citep{PhysRevLett.58.1494, PhysRevLett.58.1490, ALEXEYEV1988209}, these results played an important role in confirming the mechanism of supernova explosions~\citep{Pagliaroli:2008ur} and made many studies of neutrino properties, such as limits on the neutrino mass, neutrino lifetime, gravitational attraction of neutrinos, and so on.
In order to establish and discuss the detailed mechanisms of supernovae and refine our knowledge of neutrino properties~\citep{Horiuchi:2018ofe}, further supernova neutrino signals have been searched for about 40 yr with various detectors. 
The average energy of supernova neutrinos is about 10\,MeV and the typical time scale of neutrino emission is about 10\,s.
These supernova neutrinos are observed as event clusters in a neutrino detector.
Unfortunately, there has been no significant event clusters observed except for SN 1987A. 

Currently, the most sensitive detector is Super-Kamiokande, which has a fiducial volume of 22.5\,kt.
They reported searching for supernova neutrinos from 1996 May to 2005 October and set the upper limit on the supernova rate:  0.32\,yr$^{-1}$ within 100\,kpc~\citep{Super-Kamiokande:2007zsl}. 
Recently, a newer result was reported in \citet{Super-Kamiokande:2022dsn}.
The Large Volume Detector (LVD) and Baksan have also reported supernova neutrino searches with their long-term experimental data. 
The LVD set the supernova rate was smaller than 0.114\,yr$^{-1}$ within 25\,kpc from their data of 1992 June$\mathrm{-}$2013 December~\citep{LVD:2014uzr}. 
Baksan provided the constraint on the supernova rate, $<$ 0.070\,yr$^{-1}$, within our Galaxy from their data of 1980 June$\mathrm{-}$2018 December~\citep{NOVOSELTSEV2020102404}.
Additionally, Sudbury Neutrino Observatory~\citep{Collaboration:2010gx},  MiniBooNE~\citep{MiniBooNE:2009sim}, and the Irvine-Michigan-Brookhaven
detector~\citep{Miller:1994wq} have also searched for astronomical neutrino bursts.
From historical supernovae in astronomy, the Galactic supernova rate is predicted as 0.032$^{+0.073}_{-0.026}$\,yr$^{-1}$~\citep{Adams:2013ana}.
In other studies, its rate is predicted as $0.0163\pm 0.0046$\,yr$^{-1}$~\citep{ROZWADOWSKA2021101498}.
This value is calculated by combining various determinations of the supernova rate, which include neutrino observations.

KamLAND, an antineutrino detector using a 1\,kt liquid scintillator, also has a significant sensitivity to electron antineutrinos from supernovae owing to a low radioactive environment, the largest target volume among world liquid scintillator detectors, and a low energy threshold. 
This paper presents the results from a neutrino burst search using the KamLAND data set. 
In Section~\ref{sec:detector}, we describe the KamLAND detector and dataset employed.
The neutrino event selection criteria and background studies are summarized in Section~\ref{sec:selection}.
We present the result of the neutrino cluster search and the upper limit on the rate of neutrino clusters in Section~\ref{sec:analysis}.
In Section~\ref{sec:discussion}, we evaluate the detectable range of supernova neutrinos in KamLAND. Then we confirm that the obtained upper limit is considered as the upper limit on the supernova rate in our Galaxy.
We also discuss the Galactic star formation rate~(SFR) from the obtained supernova rate in that section. 
The Galactic SFR estimated from astronomical observations has a large uncertainty: 1--2~$M_{\odot}\,\mathrm{yr}^{-1}$~\citep{McKee_1997, Murray_2009, Robitaille_2010, Chomiuk_2011, 10.1111/j.1365-2966.2011.19095.x}. 
The constraint on the Galactic SFR from neutrino experiments presents independent and complementary information. 
A summary is given in Section~\ref{sec:summary}.

\section{KamLAND Detector AND DATASET} \label{sec:detector}
KamLAND is located 1000\,m underground from the top of Mt.~Ikenoyama in the Kamioka mine, Japan. 
The KamLAND detector consists of an inner liquid-scintillation detector (ID) and an outer water-Cherenkov detector (OD).
A 1\,kt organic ultrapure liquid scintillator is filled in a nylon/EVOH (ethylene vinyl alcohol copolymer) balloon, which has a 13\,m diameter.
The balloon is suspended in nonscintillating mineral oil contained inside of an 18\,m diameter spherical stainless steel tank.
The nonscintillating mineral oil shields the liquid scintillator from external radiations;
1325 17\,inch photomultiplier tubes (PMTs) and 554 20\,inch PMTs are mounted on the inner surface of the tank.
ID consists of these components described above.
The outside of the tank is OD with 140 20\,inch PMTs for muon vetoes.  
Absolute time accuracy in KamLAND is less than $\mathcal{O}(100)\,\mathrm{\mu s}$. 
Details of the detector are described in \citet{Suzuki2014}.

KamLAND began the data taking in 2002 March.
To eliminate radioactive impurities in the liquid scintillator, two distillation campaigns had been conducted from 2007 March to August and from 2008 July to 2009 February.
From 2011 August to 2015 September, a 3.08\,m diameter inner balloon had been installed at the center of the detector for the KamLAND-Zen~400 experiment, which searches for neutrinoless double-beta decay of $^{136}\mathrm{Xe}$ using a xenon-loaded liquid scintillator~\citep{Zen400}.
This inner balloon was extracted at 2015 December.
From 2016 January to June, the OD system had been refurbished to replace 225 PMTs reused from the Kamiokande experiment with 140 new PMTs including high-quantum efficiency ones~\citep{Ozaki:2016fmr}.
From 2018 May, a 3.80\,m diameter inner balloon has been installed for the KamLAND-Zen~800 experiment~\citep{zencollaboration2021nylon, KamLAND-Zen:2022tow}.
We use all the data from 2002 March 9 to 2020 April 25 except for minor noisy periods during the distillation campaigns, inner-balloon installation, and uninstallation works, and OD refurbishment works.

\section{Selection Criteria} \label{sec:selection}

We focus on electron antineutrinos ($\bar{\nu}_e$'s) detected via the inverse beta decay (IBD) interaction ($\bar{\nu}_e + p \rightarrow e^+ + n$) for this neutrino burst search.
The incident neutrino energy and event vertex are reconstructed based on the timing and charge distributions of the PMT waveforms.
IBD events are detected by the delayed coincidence (DC) method, which tags a sequential pair of a prompt (positron and annihilation gamma rays) and delayed (neutron capture gamma rays) events with space and time correlations.

We select the prompt energy range of $0.9\mathrm{-}100\,\mathrm{MeV}$ corresponding to $1.8\mathrm{-}111\,\mathrm{MeV}$ of the neutrino energy.
The delayed event is a thermal neutron capture gamma-ray on a proton or carbon.
The delayed energy range is required to be $1.8\mathrm{-}2.6\,\mathrm{MeV}$ for the neutron capture on a proton and $4.4\mathrm{-}5.5\,\mathrm{MeV}$ on a $^{12}\mathrm{C}$.
The time difference between the prompt event and delayed event is required to be in $0.5$ to $1000\,\mathrm{\mu s}$, and the vertex distance is to be within $160\,\mathrm{cm}$.
We use a spherical fiducial volume with a 600\,cm radius from the center of ID, for both prompt and delayed events, corresponding to $\mathcal{N}_{\mathrm{target}} = (5.98 \pm 0.13) \times 10^{31}$ of target protons.
During the KamLAND-Zen phases, we additionally veto delayed events in the inner-balloon region.
The cut region is a 250\,cm radius spherical volume from the center and a 250\,cm radius cylinder volume in the upper half of the detector.

In order to suppress accidental DC contamination, we use the maximum likelihood selection, which depends on the energy and vertex of the prompt and delayed events, and the time difference and distance.
At the energy above $4\,\mathrm{MeV}$, selection efficiencies ($\epsilon_{\mathrm{eff}}(E)$) for IBD events are almost constant, and whose values are $\sim$77\% in the KamLAND-Zen periods with the inner-balloon volume cut and $\sim$94\% in other periods with the full fiducial volume. As a result of the likelihood selection, these efficiencies are decreased due to radioactive impurities at the energy below $4\,\mathrm{MeV}$.

To reduce multiple neutron capture gamma-rays induced by cosmic-ray muons, they are tagged by Cherenkov radiation in OD and scintillation light in ID.
We apply a 2\,ms veto for the whole detector volume and a 2\,s veto for the cylindrical volume along the muon track after muons~\citep{SolarAntivKamLAND2012}.
Due to muon vetoes, the typical livetime ratio is $\sim$88\% and the total livetime in our report is 5011.51\,days.

\begin{figure}[htbp]
    \centering
    \includegraphics[width=0.8\linewidth]{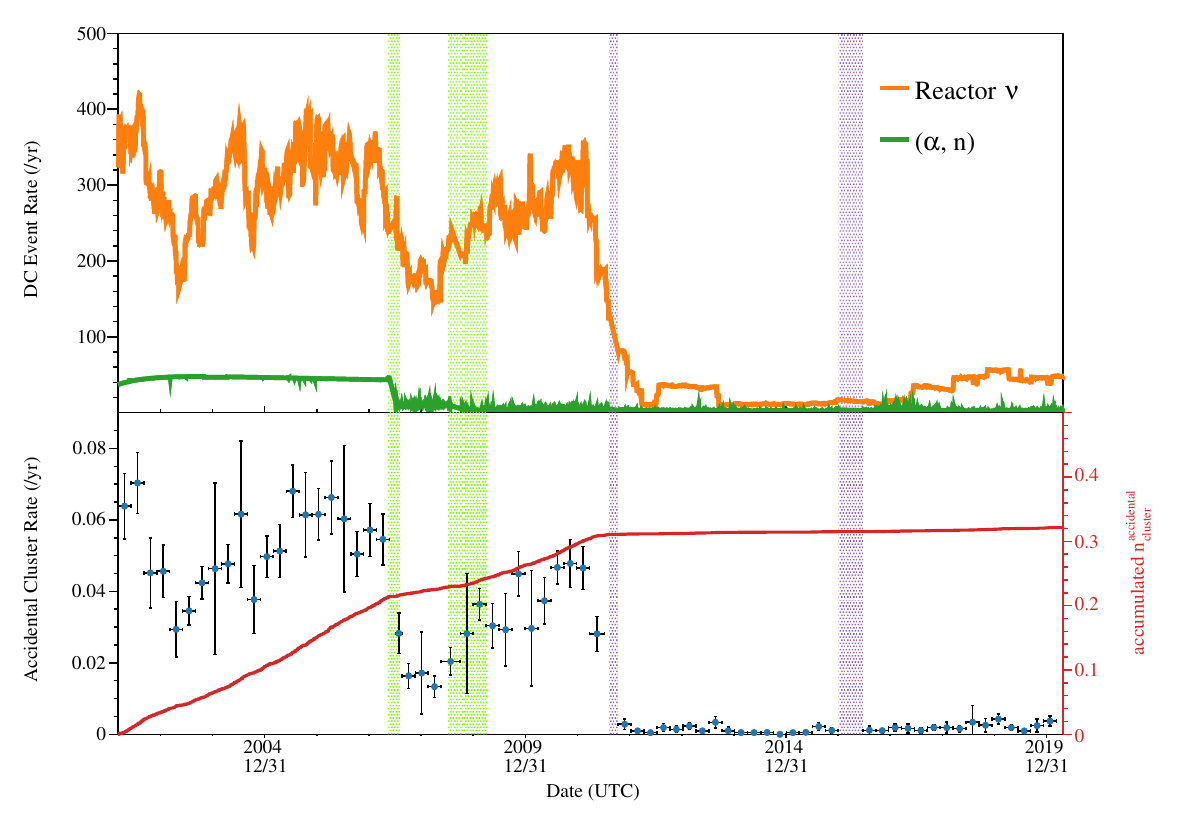}
    \caption{DC event rate (top) and accidental cluster rate, and accumulation of the number of accidental cluster events (bottom) as a function of date.
    In the top panel, the orange line shows the expected DC event rate of reactor -$\bar{\nu}_e$ and the green line shows that of $^{13}\mathrm{C}(\alpha, n)^{16}\mathrm{O}$ interaction.
    In the bottom panel, blue dots represent averaged accidental cluster rates.
    Error bars show the period and the statistic error comes from the weighted average based on the livetime.
    The red curve means the accumulated number of accidental clusters.
    Vertical light green bands are the distillation campaigns. 
    The vertical violet band in 2011 is the balloon-installation campaign for KamLAND-Zen~400, and the band in 2016 is the OD refurbishment period.
    For another balloon installation for KamLAND-Zen~800, we have a too short deadtime period to be shown here.
    }
    \label{fig:livetime}
\end{figure}

To detect supernova neutrinos with the KamLAND data, a cluster of DC events is required: two DC events within a 10\,s window.
The very low DC rate enables us to use the minimum cluster condition (two DC events).
An accidental DC event cluster is a background for supernova neutrinos.
Based on the DC event studies, we evaluate the number of accidental cluster events~($\equiv n_{\mathrm{cluster}}^{\mathrm{accidental}}$) via Monte-Carlo (MC) simulation.
Accidental cluster rates and its accumulation are shown in Figure~\ref{fig:livetime}.
Before distillation campaigns, the DC event is mainly caused by reactor -$\bar{\nu}_e$, and $^{13}\mathrm{C}(\alpha, n)^{16}\mathrm{O}$ interaction has a secondary contribution~\citep{PhysRevLett.100.221803}.
After distillation campaigns, the DC event by $^{13}\mathrm{C}(\alpha, n)^{16}\mathrm{O}$ interaction decreases but by reactor -$\bar{\nu}_e$ is still a dominant component~\citep{PhysRevD.83.052002}.
Japanese reactors were shut down after the Great East Japan Earthquake in 2011.
In the reactor-off phase~\citep{PhysRevD.88.033001}, the rate of reactor -$\bar{\nu}_e$ events is decreased.
These trends of DC event rate are also shown in Figure~\ref{fig:livetime}.
In an energy region above 10\,MeV, fast neutrons and atmospheric neutrinos have dominant contribution on the number of DC events~\citep{SolarAntivKamLAND2021}.
Typically, the DC candidate rate is $\sim1\,\mathrm{day}^{-1}$ in the reactor-on phase and $\order{0.1}\,\mathrm{day}^{-1}$ in the reactor-off phase.
Usually, we call a KamLAND data set as a ``run,'' which consists of typically 24\,hr data.
The number of expected DC events is estimated in run by run and the number of accidental clusters is calculated from that.
Taking into account that time differences between each run are more than 10\,s, there should be no accidental clusters that occur across two runs.
As a result, the accumulated number of accidental clusters is $n_{\mathrm{cluster}}^{\mathrm{accidental}}= 0.32^{+0.02}_{-0.04}$\,clusters, and this rate is $0.023\,\mathrm{cluster}\,\mathrm{yr}^{-1}$ on average.
The error of $n_{\mathrm{cluster}}^{\mathrm{accidental}}$ includes systematic uncertainties of the number of expected DC events by reactor -$\bar{\nu}_e$, atmospheric neutrino, and fast neutron.

\section{Analysis and Result} \label{sec:analysis}
\begin{figure}[htbp]
    \centering
    \includegraphics[width=0.8\linewidth]{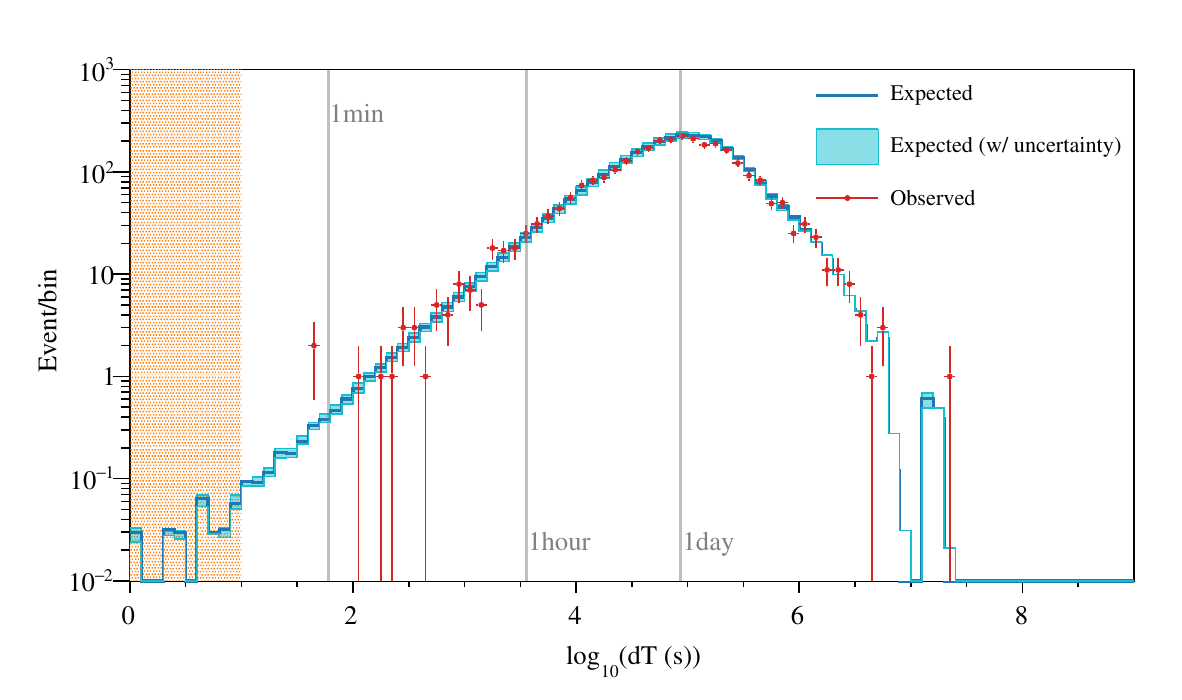}
    \caption{
    Distribution of the time differences between DC candidates.
    Red points represent observed events.
    The blue histogram represents expected events and the light blue band shows systematic uncertainties as described in the Section~\ref{sec:selection}.
    The orange vertical hatched region corresponds to a time difference less than 10\,s, for a  supernova burst neutrino observation.
    Gray vertical lines represent time differences between 1 minute, 1 hr, and 1 day for each other.
    }
    \label{fig:dT}
\end{figure}

We have selected DC candidates and searched for neutrino clusters from all the data sets of KamLAND.
The Figure~\ref{fig:dT} shows time differences among the DC candidates. 
We find no clusters within the 10\,s time window ($n_{\mathrm{cluster}} = 0$).
The closest time difference of DC candidates is 41\,s, observed at 2003 September 24 15:26:15 (UTC) and 2003 September 24 15:26:56 (UTC).
The second and third closest time differences are 42 and 103\,s;
the expected distribution of the time difference assuming no supernova bursts show the agreement with real time differences, and the total number of estimated accidental clusters is $n_{\mathrm{cluster}}^{\mathrm{accidental}} = 0.32$, which is referred to in Section~\ref{sec:selection}.
Following the Feldman$\mathrm{-}$Cousins approach~\citep{FeldmanCousinsMethod}, we set the upper limit as $n_{\mathrm{cluster}} < 2.1$
with a 90\% confidence level (CL) in the livetime of 5011.51\,days.
From this result, we can derive the upper limit on the cluster rate as $R_{\mathrm {\,cluster}} < 0.15\,\mathrm{yr}^{-1}$ with a 90\% CL.

\section{Discussion} \label{sec:discussion}
\begin{figure}
    \centering
    \includegraphics[width=0.7\linewidth]{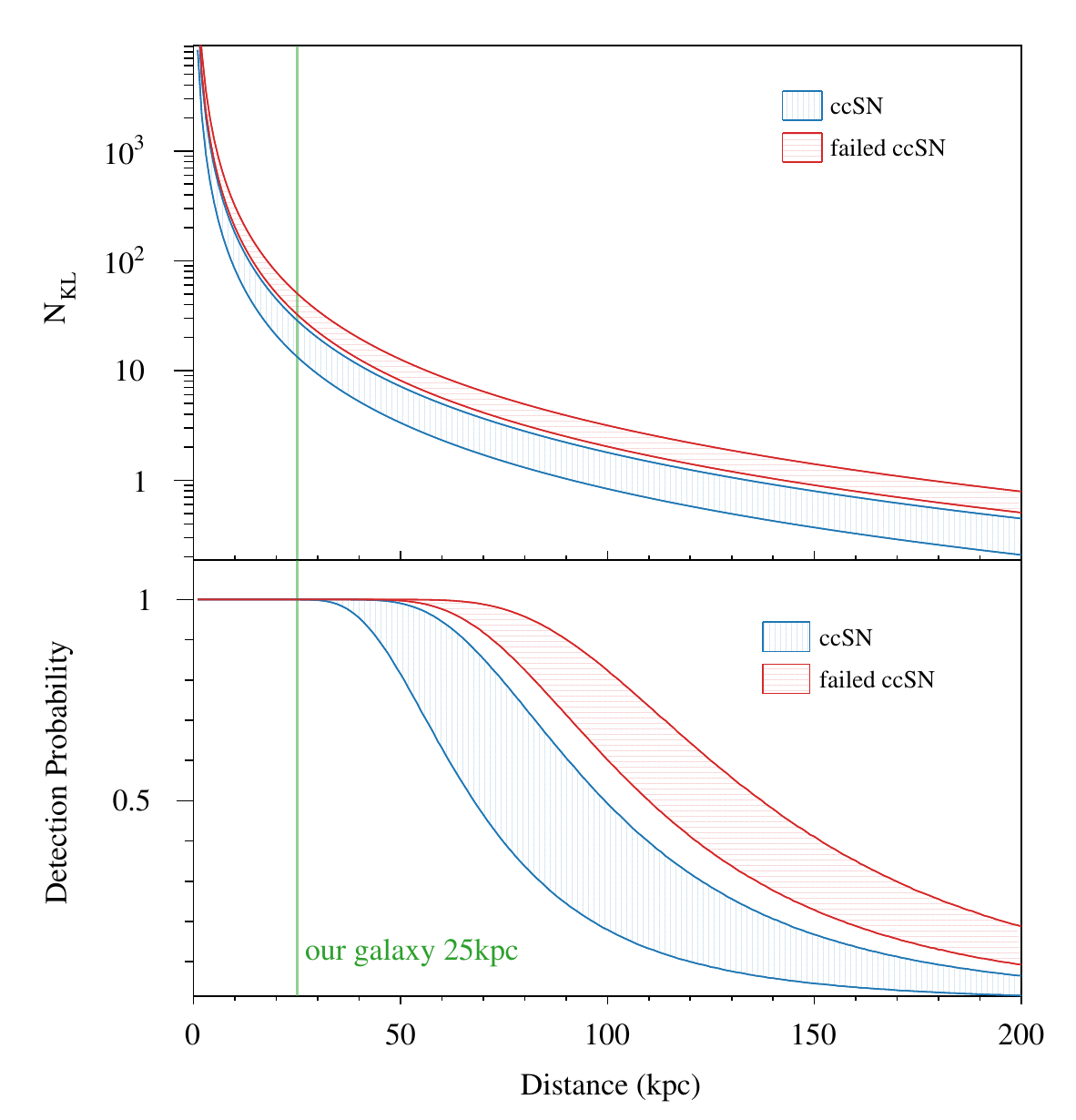}
    \caption{The expected number of observed supernova events $N_{\mathrm{KL}} (r)$ (upper figure) and detection probabilities (lower figure). Both curves are calculated employing the Nakazato model~\citep{Nakazato_2013}. The blue band indicates the range for ccSNe, and the red band represents failed ccSNe. Green vertical line represents the scale of our Galaxy.}
    \label{fig:detprob}
\end{figure}

In the previous section, we found no clusters and set the upper limit on the number of observed clusters. 
In this section, we consider that the clusters are originated from supernovae, and evaluate the detectable range. 
We use the number of neutrinos emitted from core-collapse supernovae (ccSNe) and failed ccSNe, which result in black hole formations, derived from the Nakazato model~\citep{Nakazato_2013}.
This model provides 20\,s supernova neutrino data and is parameterized by an initial mass $M_{\mathrm{init}}$, metallicity $Z$, and shock revival time $t_{\mathrm{revive}}$.
The electron antineutrino flux arriving at the Earth, $F_{\bar{e}\mathrm{Earth}} (r)$, is written as~\citep{Dighe:1999bi}
\begin{equation}
    \label{eq:finalflux}
    F_{\bar{e}\mathrm{Earth}} (r,E,t) = \frac{1}{4\pi r^2}\qty(\bar{p} \frac{\dd{^2 N_{\bar{e}0}}}{\dd{E}\dd{t}} + (1-\bar{p})\frac{\dd{^2 N_{x0}}}{\dd{E}\dd{t}}),
\end{equation}
where $r$ is the distance from the Earth to a supernova, $\dd{^2N_{\bar{e}0}}/\dd{E}\dd{t}$ is the number of emitted electron antineutrinos per MeV per second, $\dd{^2N_{x0}}/\dd{E}\dd{t}$ is that of the antineutrinos $\bar{\nu_{\mu}}$, $\bar{\nu_{\tau}}$ and $\bar{p}$ is the survival probability, which is 0.665 (normal mass ordering) or 0.0216 (inverted mass ordering) with $\sin^2{\theta_{12}}=0.320$ and $\sin^2{\theta_{13}}=0.0216$~\citep{DESALAS2018633}.
We multiply $F_{\bar{e}\mathrm{Earth}} (r,E,t)$ by the cross section of the IBD $\sigma_{\mathrm{IBD}}(E)$~\citep{STRUMIA200342}, selection efficiency $\epsilon_{\mathrm{eff}}(E)$, livetime ratio $\eta_{\mathrm{\,livetime}}$, and number of target protons in KamLAND $\mathcal{N_{\mathrm{target}}}$, then integrate with the time of 20\,s and the neutrino energy of $1.8\mathrm{-}111\,\mathrm{MeV}$.
The expected number of observed supernova events $N_{\mathrm{KL}} (r)$ at KamLAND as a function of $r$ is estimated from 
\begin{equation}
    \label{eq:NofKL}
    N_{\mathrm{KL}} (r) = \eta_{\mathrm{\,livetime}} \,\mathcal{N_{\mathrm{target}}} \int \dd{t}\dd{E} F_{\bar{e}\mathrm{Earth}} (r,E,t) \sigma_{\mathrm{IBD}}(E) \epsilon_{\mathrm{eff}}(E).
\end{equation}
Actually, observed events in KamLAND follow a Poisson distribution with mean $N_{\mathrm{KL}}$.
To reproduce the time distribution of supernova neutrinos in KamLAND, we make the probability density function (PDF) of them.
We carry out MC simulation based on the above Poisson distribution and PDF, then search for a neutrino cluster that requires two DC events within a 10\,s window.
$N_{\mathrm{KL}}$ yields $10\mathrm{-}30$ events for a ccSN within 10\,s in case of $r=25\,\mathrm{kpc}$.
Here, the number of accidentally contaminated DC events in the supernova neutrino burst is negligibly small ($\order{10^{-5}}\,\mathrm{events}$) within the distance we expect $N_{\mathrm{KL}}>1$.
These calculations provide detection probabilities as a function of distance as shown in Figure~\ref{fig:detprob}.
We use all available parameter combinations in this estimation, thus blue and red bands include model and neutrino mass ordering uncertainties.
KamLAND has a 95\% probability to the supernova neutrino burst detection with the distance for $r \leq 40\mathrm{-}59\,\mathrm{kpc}$ and $r \leq 65\mathrm{-}81\,\mathrm{kpc}$ for the ccSN and failed ccSN, respectively. 
In either case, our Galaxy ($r \lesssim$ 25\,kpc) is covered with a $\geq$~99\%  detection probability.
Consequently, this result gives an upper limit on the supernova rate within our Galaxy, which includes ccSN rate and failed ccSN rate, $R^{\mathrm{\,gal}}_{\mathrm {\,SN}} < 0.15\,\mathrm{yr}^{-1}$
(90\% CL) assuming that the SN rate on the Large Magellanic Cloud and Small Magellanic Cloud are much smaller than the Galactic SN rate~\citep{Tammann:1994ev}.

\begin{figure}[htbp]
    \centering
    \includegraphics[width=0.8\linewidth]{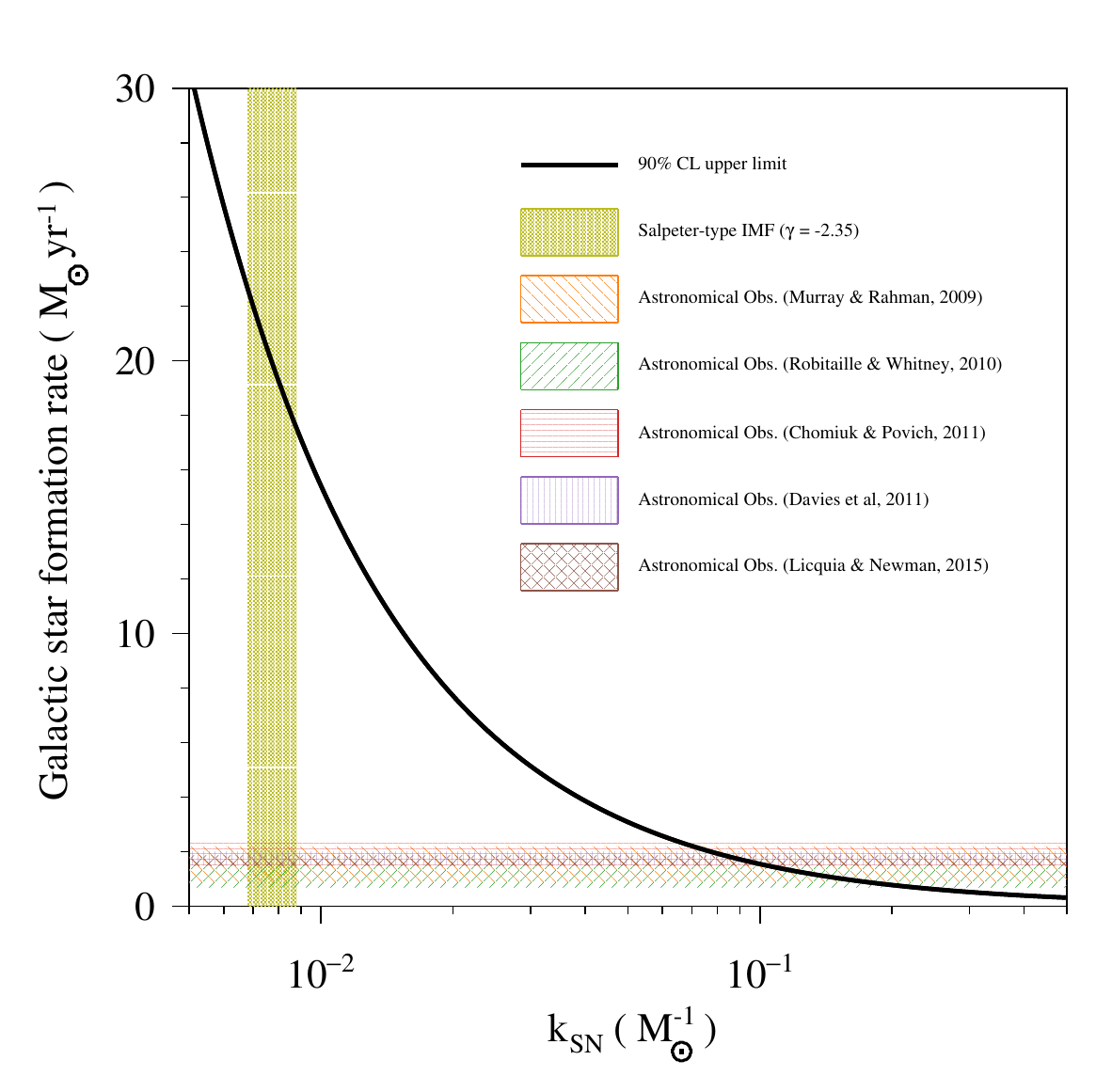}
    \caption{Upper limits on the parameter space between $k_{\mathrm{SN}}$ and $\psi^{\mathrm{\,gal}}_{\mathrm{\,SFR}}$ with 90\% CL. 
    The upper right region is disfavored, and the lower left region is favored.
    The yellow vertical band is an expected range based on the Salpeter-type IMF. 
    The horizontal bands are the allowed regions from astronomical observations.
    }
    \label{fig:sfr_vs_kcc}
\end{figure}

Next, we try to convert the Galactic SN rate into a constraint on the Galactic SFR. According to \citet{Horiuchi:2011zz} and \citet{Botticella2012}, a cosmological SN rate is linked to the cosmological SFR as
\begin{equation}
    \label{eq:snr_to_sfr}
    R_{\mathrm{\,SN}}\,(z) =
    \frac
    {\int_{m_{\mathrm{l}}^{\mathrm{SN}}}^{m_{\mathrm{u}}^{\mathrm{SN}}} \phi_{\mathrm{IMF}}\,(m) \dd{m}}
    {\int_{m_{\mathrm{l}}}^{m_{\mathrm{u}}} m \phi_{\mathrm{IMF}}\,(m) \dd{m}}
    \, \psi_{\mathrm{SFR}}\,(z) 
    \equiv 
    k_{\mathrm{\,SN}}\,\psi_{\mathrm{SFR}}\,(z),
\end{equation}
where $R_{\mathrm{\,SN}}\,(z)$ is the SN rate, $\psi_{\mathrm{SFR}}\,(z)$ is the SFR as a function of redshift~$(z)$, $\phi_{\mathrm{IMF}}$ is the initial mass function (IMF), $m$ is the mass of a star, ($m_l$--$m_u$) is the mass range of IMF, ($m_{\mathrm{l}}^{\mathrm{SN}}$--$m_{\mathrm{u}}^{\mathrm{SN}}$) is the mass range of SN stars, and $k_{\mathrm{\,SN}}$ is the scaling factor between $R_{\mathrm{\,SN}}$ and $\psi_{\mathrm{SFR}}$ by the number fraction of stars per unit mass.
We apply this relationship for the Galactic SN rate and averaged Galactic SFR; therefore, $R^{\mathrm{\,gal}}_{\mathrm{\,SN}}=k_{\mathrm{\,SN}}\,\psi^{\mathrm{\,gal}}_{\mathrm{SFR}}$.
We show the constraint on $\psi^{\mathrm{\,gal}}_{\mathrm{SFR}}$ and $k_{\mathrm{\,SN}}$ with
90\% CL as shown in Figure~\ref{fig:sfr_vs_kcc}.
The yellow vertical band is an expected range of $k_{\mathrm{\,SN}}$ = (0.0068--0.0088)\,$M_{\odot}^{-1}$ assuming the (modified) Salpeter-type IMF~\citep{SalpeterIMF, Horiuchi:2011zz, doi:10.1146/annurev-astro-081811-125615} which contains uncertainties come from astronomical observations and models.
The lowest value of $k_{\mathrm{\,SN}}$ is calculated from the IMF with a power index of the mass $\gamma = -2.35$ in the mass range of 0.1--100\,$M_{\odot}$.
On the other hand, the highest value of $k_{\mathrm{\,SN}}$ is obtained by combining the IMF with $\gamma = -1.5$ in the mass range of 0.1--0.5\,$M_{\odot}$ and the one with $\gamma = -2.35$ in the mass range of 0.5--100\,$M_{\odot}$.
In either case, the mass range of SN stars is set in 8--40\,$M_{\odot}$.
The colored horizontal lines and bands correspond to an allowed $\psi^{\mathrm{\,gal}}_{\mathrm{SFR}}$ range reproduced from the astronomical observation as to be (1--2)\,$M_{\odot}\,\mathrm{yr}^{-1}$~\citep{Murray_2009, Robitaille_2010, Chomiuk_2011, 10.1111/j.1365-2966.2011.19095.x, Licquia_2015}.
Our result thus provides the upper limit on the SFR as
$\psi^{\mathrm{\,gal}}_{\mathrm{SFR}}$\,$<$\,(17.5--22.7)\,$M_{\odot}\,\mathrm{yr}^{-1}$ with 90\% CL assuming the Salpeter-type IMF within our Galaxy.
This result disfavors a large SFR in our Galaxy and is consistent with the constraints from the astronomical observations.

\section{Summary} \label{sec:summary}
We have presented the results of the search for supernova neutrino clustering events in the KamLAND data set from 2002 March 9 to 2020 April 25.

The expected number of accidental neutrino clusters is 0.32 and the observed number of neutrino clusters was zero.
We set a 90\% CL upper limit on the supernova rate in our Galaxy as $R^{\mathrm{\,gal}}_{\mathrm{SN}} < 0.15\,\mathrm{yr}^{-1}$.
This result corresponded to a 90\% CL upper limit on the Galactic SFR of $\psi^{\mathrm{\,gal}}_{\mathrm{SFR}}$\,$<$\,(17.5--22.7)\,$M_{\odot}\,\mathrm{yr}^{-1}$.

Further studies of supernova neutrinos will provide important insights not only into the explosion mechanism and the nature of neutrinos, but also into the SFR of the Milky Way.

\begin{acknowledgments}
The KamLAND experiment is supported by 
JSPS KAKENHI grants 
19H05803; 
the World Premier International Research Center Initiative (WPI Initiative), MEXT, Japan; 
Netherlands Organization for Scientific Research (NWO); 
and under the U.S. Department of Energy (DOE) Contract 
No.~DE-AC02-05CH11231,
the National Science Foundation (NSF) No.~NSF-1914418, 
NSF-2012964, 
NSF-2110720, 
as well as other DOE and NSF grants to individual institutions.  
The Kamioka Mining and Smelting Company has provided services for activities in the mine.  
We acknowledge the support of NII for SINET. 
We also thank M.\,Tanaka for helpful discussion.
This work is partly supported by 
the Graduate Program on Physics for the Universe (GP-PU).
\end{acknowledgments}


\bibliography{main.bib}{}
\bibliographystyle{aasjournal}

\end{document}